# Reflections of quantum educators on strategies to diversify the second quantum revolution

Apekshya Ghimire* and Chandralekha Singh
*Department of Physics and Astronomy, University of Pittsburgh, 3941 O'Hara St, Pittsburgh, PA, 15260*

*apg61@pitt.edu

## Introduction

The second quantum revolution focuses on quantum information science and technology (QIST). It promises to bring about unprecedented improvements in computing, communication and sensing due to our ability to exquisitely control and manipulate quantum systems [1, 2]. This fast-growing field presents equally unprecedented opportunities and challenges for preparing students to be future leaders in this area [3, 4].

One major challenge pertains to how to diversify the second quantum revolution. The first quantum revolution, which began nearly a century ago and laid the foundation for the technological breakthroughs of the mid-20th century, reflects the broader lack of diversity that characterized the scientific landscape of that era. Since the second quantum revolution is in its infancy, it is critical to contemplate these issues to ensure that people from historically marginalized groups in physics and related fields can contribute equitably [5, 6].

This paper focuses on reflections and suggestions of five college quantum educators from four different institutions (two from the same institution) regarding what can be done to diversify the second quantum revolution. They are leading QIST researchers, and very passionate about improving quantum education. The educators were also asked about their thoughts on whether the interdisciplinary nature of the field, in which nobody can claim to be an expert in all aspects of QIST, may make it easier to create a better culture from the beginning, supportive of equitable participation of diverse groups unlike physics. This is because disciplines such as physics have an ingrained inequitable culture based on brilliance attribution [7] that is a major impediment to diversity, equity and inclusion. Educators were interviewed on Zoom using a semi-structured think-aloud protocol about various issues related to QIST education including those pertaining to how to diversify the second quantum revolution discussed here.

Below, we discuss some excerpts that exemplify interviewed quantum educators' reflections and suggestions in their own words. They are insightful and can help other educators adapt and implement strategies to diversify QIST. We mention below that one of the educators was a woman because her gender identity may be useful to understand her point of view.

## Quantum educators in their own words

Some educators emphasized the importance of teaching quantum concepts early, e.g., in high school and early in college to students from diverse backgrounds. This exposure is important because targeting diverse high schools has the potential to introduce exciting quantum concepts and careers to students from diverse backgrounds considering diversity is severely lacking among college STEM majors. Early introduction is also valuable because learning quantum concepts early may make the subject matter less mysterious. Students may also better learn to discern the contexts in which classical and quantum concepts are applicable. This is similar to the facility children have in discerning which language to use in which contexts and switching between them as needed depending upon contexts, when they simultaneously learn two different languages early. Emphasis on potential QIST applications focusing on solving societal problems that resonate with students and discussing career pathways with students can be useful.

For example, one male educator emphasized the importance of teaching quantum and relevant mathematical concepts early saying, "If I had the magic wand that could reform, you know, high school curriculum, I would probably change a few [parts] of the mathematics teaching. Like, I think, data science and statistics at this point should have at least as much space as calculus does. Well, once you understand statistics…then quantum is a small step from there and [can even be taught with] linear algebra. I would teach…coding, programming, statistics... And once you've got that, you can teach any quantum you want…even in high school…My course, that I am teaching second year [college students] here, with minor modifications, I could teach in high school."

He also emphasized the importance of helping students from diverse backgrounds develop confidence along with competence from early on while learning QIST concepts saying, "in every field, no matter how developed or under-developed [like QIST] it is, you will find people who have a lot of confidence, you know, becoming the most visible ones…It's something that students should learn a little in high school…".

Another male educator who is passionate about early teaching as well as early research experiences related to QIST for first- and second-year college students in science and engineering reflected on the importance of supporting all students in foundational courses especially those not particularly interested in physics saying, "let's say that we [with physics background] are teaching a college course on foundations of quantum computing…we found some advisors who said to students in electrical engineering or computer science that this might be a good thing if they were thinking about…going into the workforce in this area. Now when they landed in those courses…they're like, oh, this sounds too much like physics to me. So what can you do?...We want all these students if they're going to take an intro course…to succeed…there should be something for everybody...They shouldn't be punished for not being a physicist". He acknowledged that there are many ways of structuring interdisciplinary QIST courses for early learners, but felt educators need to emphasize applications in quantum computing, communication and sensing and how they could impact society through examples that resonate with students from diverse backgrounds as well as discuss career pathways. He also noted that while these early courses should mainly focus on learners with diverse background preparations and interests, students can be provided enrichment options through projects/special modules as part of the course that dive more deeply into the sub-field of QIST, e.g., engineering or computer science, they may be more interested in. This educator also noted leading a first experiences in quantum research program for first-year college students currently that he hopes will play a role in diversifying QIST.

He also stated that diversifying issues are not only in QIST but in many STEM disciplines including physics stating, "simply acknowledging this…and trying to make sure that people have support that they need throughout their…education to be able to…walk into a room where people don't look like them and succeed" is important. He did not think that QIST being a new field makes it easier to diversify than the sub-fields such as physics, "I do think that there's in some sense…we could just like start over afresh, because this [QIST] is a new field, and we don't carry over the baggage…I'm not sure that it's succeeding so far". He felt that the current diversity in QIST is dismal saying, "I don't see things as automatically getting fixed just because we rebranded. And now it's quantum 2.0. I think that it's…very important to push back against these…And I think that the earlier the positive experiences people have in doing that [learning quantum concepts] successfully, the better the outcome could be, that you have to…realize that, you know, imposter syndrome is like, almost everybody feels that…has that kind of feeling at some stage or another and that you…can embrace it and say, well, I'm new to this field, but I still belong here…[do] some type of self-affirmation".

He also reflected on the highly interdisciplinary nature of QIST playing a role in sense of belonging and emphasized the role of educators and researchers in contemplating ways to integrate students in QIST, who are trained differently. He felt that since the preparation of students is siloed in physics, engineering, computer science etc., some students may feel uncomfortable being part of this new interdisciplinary field of QIST saying, "what's interesting about the quantum is…there are dimensions to [it] where people feel like they don't belong…they have to do with their professional upbringing, right? So, somebody can feel that they don't belong…because they weren't trained in a certain way and I think that it's important to address all of these issues, right? Whatever it is that is inhibiting people from learning and doing [QIST]".

A female educator reflected on these issues saying, "I think, in general…physics has been very sort of depressingly bad…I'd say that quantum does not have good representation, you know it's not so inclusive…".

She also felt that unless the culture of each subfield, e.g., physics, that makes up QIST changes, she disagreed with "this idea that there are these different fields with somewhat different cultures and maybe that could provide an avenue to improving [QIST]" emphasizing that a major barrier to diversity is, "this thing of like, if you get a guy, and he knows, you know, half of something, [he'll say] oh, yeah, I know that. And you get a woman, and she knows more, [she'll say] oh, I don't know…everybody's faking it. But somehow people's internalization of what they know relative to everyone else is very culturally dependent". She felt that unless we can do something to eradicate these types of dichotomies that are systemic in the culture, e.g., overly confident men making statements bragging about more than what they know and women undermining themselves, the new field of QIST will continue to be intimidating for many women and other underrepresented people similar to physics and they would decide not to be part of a field with such an inequitable culture.

She gave several concrete examples of men being overconfident and acting as if they were pinnacles of brilliance based upon her own interactions with them in physics. She exemplified specific cases in which they knew significantly less than her but pretended to be on top of everything and took the opportunity to promote themselves. She reiterated that she is worried that this kind of inequitable culture that pervades physics will seep into this new field of QIST since the same people who have dominated the space in physics will continue to do so in QIST, "I'm saying just like having that confidence of, you know, I belong here...[I disagree that] one of the great things about a field that's as broad as this one [QIST] is that nobody knows everything. But you don't feel that way [if you are an underrepresented person such as a woman], you feel like everybody knows it but you [because of the way many men portray themselves]". She felt that the only way to diversify QIST is to change the systemic cultural issues in physics and other related disciplines that make up QIST in which brilliance is over-emphasized and many individuals from the dominant group are pretending and portraying to know more than they do, thereby intimidating others and even driving them out of the discipline.

Another male educator strongly believed that early outreach to students from diverse backgrounds in various settings can play a critical role in diversifying QIST saying, "we've got a massive issue with diversity in physics and…in electrical engineering and other engineering disciplines…in computer science…and I think we, as physicists, have to be very active in going out to the general public".

He also emphasized the importance of increasing the visibility and impact of underrepresented people in QIST saying, "And I think that it's important to emphasize and underline the contributions of those role models and to show to people that this is a field for everyone". He thought it was important that "when we organize events, we represent the diversity of people who are involved…there are people from a lot of minority ethnic groups involved in this area. There are a lot of women involved in this area really doing internationally leading research". To diversify QIST, he stressed the need to change the culture of physics and related disciplines comprising QIST that currently favors the dominant group, "we need to keep looking at our own research teams, our ways of interacting with people. We need to look at our recruiting practices. We need to look at the ways in which we organize meetings, and we need to be constantly questioning whether we have tried to understand all of this best practice from what's been done in this field around the world, making sure that we implement it in our own teams and make sure that we essentially spread the message and take leadership roles in saying that diversity, equity, and inclusion, these things are extremely important to us, and extremely important to the health of the field, going forth".

He added, "we lose out still a lot. A lot of very talented physicists choose to go into other areas…they have bad experiences with a workplace culture...I also think that when you get a diverse range of people, people with different ethnic backgrounds, people of different nationalities, people from different gender backgrounds, people from…all different ways you can be, I find that teams tend to work much better. They tend to be more creative because people coming in with different ideas, different ways [of] thinking lead to different types of creativity, but also the culture of teams change…most inefficient committees I've ever been on, we're essentially all made up of white men and the most productive committees I was on, we're typically always characterized by having a very diverse range of people. And I think for the health of our field, it's important that we take leadership roles and that we do everything we can in this area".

Another male educator, who acknowledged that the physics culture is a detriment to diversifying, was concerned that this burgeoning QIST field can morph into a non-diverse field like physics saying, "physics, of course, has just long problems with underrepresented minorities, which because quantum technology is physics led, is a problem that [can] propagate into that". However, he expressed optimism that "attitude in physics might not hopefully propagate to wider quantum technology, and so might be a way to kind of get…minorities in…so there's that sense of optimism there that with the invention of a new field [QIST], you can invent a new culture hopefully, free from some of the mistakes of physics".  He also felt that whatever strategies are currently being used to increase diversity, equity and inclusion in physics should be employed in QIST since they may be useful for QIST as well.

**Discussion and Summary**

The interdisciplinary field of QIST is fast-growing with major implications for the future workforce. This paper focuses on reflections and suggestions from five college quantum educators, who are passionate about teaching quantum concepts, regarding how to diversify the second quantum revolution. Their suggestions can be invaluable for other educators interested in similar pursuits and can be summarized as follows:

- Focus on changing the culture of disciplines such as physics since the inequitable culture in these sub-disciplines comprising QIST can impact outcomes in QIST, which is interdisciplinary.
- Acknowledge the need for diversity, equity, and inclusion in QIST and support students from diverse backgrounds as appropriate.
- Conduct quantum outreach in both formal (K-12) and informal settings to get students from diverse backgrounds aware of and interested in QIST concepts and career opportunities.
- Teach quantum concepts early, e.g., in high school and early in college to students from diverse backgrounds.
    - More focus in high school on statistics, data science, and programming may be helpful for introducing quantum early.
    - Focus on developing student confidence along with competence.
    - Focus on not teaching early QIST courses as physics courses
    - Tailor courses to benefit diverse backgrounds and interests emphasizing applications in quantum computing, communication and sensing that have the potential to solve major societal problems as well as career pathways.
- Give students early experience in QIST research.

- Increase the number of role models in QIST.
- Increase the visibility and impact of existing underrepresented QIST researchers, e.g., by including them as invited speakers and on committees that make major decisions.
- Try similar approaches to those that have proved useful for diversifying physics.

Based upon these suggestions, the involvement of early learners, e.g., in high school, college and via outreach is important for ensuring that students from diverse backgrounds can learn about exciting quantum concepts and careers particularly because college STEM majors lack diversity. There is a need to ensure that students develop an appropriate level of confidence and competence commensurate with these courses. Early exposure at the appropriate level can make quantum concepts less spooky and ensure that students recognize the differences between situations in which classical and quantum frameworks are useful to understand physical phenomena. It would help to emphasize QIST applications and how they can help solve major societal problems such as making nitrogen fixation more energy efficient for agricultural innovation and drug discovery. In addition, discussion of career pathways and the relevance of QIST to national security can be valuable. Giving students opportunities for early experiences in QIST research as appropriate, commensurate with their prior preparation, can also get students excited about pursuing QIST related careers.

Educators have begun to develop educational tools for early quantum learners. For example, tools have been developed to visualize QIST concepts [8-10], explore physical realizations of qubits [11], as well as approaches that allow for kinesthetic [12] and game-based learning [13-15] to understand entanglement, decoherence, Bell's inequality and quantum cryptography. Educators can engage students by using other interactive approaches that use "compare and contrast" activities involving quantum and classical concepts [16, 17], diagrammatic tools [18], evidence-based teaching-learning sequences [19] and optics experiments [20, 21] for early learners to get students excited about QIST concepts [22].

Some organizations such as National Q-12 partnership [23] and Quantum for All [24, 25] are providing a community to K-12 educators interested in incorporating quantum concepts early. For example, National Q-12 partnership [23] is promoting and supporting quantum education by connecting those with interest in K-12 quantum education with each other as well as hosting a framework for K-12 quantum education that can be used to develop QIST curricula. They also host a variety of QuanTime activities, which are ready-to-use activities, that K-12 educators can use in their classes without having prior experience with teaching QIST concepts. For example, in 2024 year, there were 27 activities available to the K-12 teachers [23] and teachers were encouraged to devote at least one class period around the World Quantum Day on April 14 on one of the activities they selected. Quantum for All [24, 25] has been a leader in providing an inclusive community for K-12 educators interested in incorporating quantum education in their classes in a conceptual manner. They provide intensive professional development activities to K-12 educators to ensure that they feel like they are part of a community with similar goals and feel confident infusing quantum concepts into their classes [24, 25].

Other organizations such as Womanium [26] are for all levels (high school through graduate school and postdoc). Womanium is founded and run by women, who are QISE researchers and educators, and it is working on bringing in students of all genders, nationalities, and levels to the QIST field[26]. Qubitbyqubit [27] and IBM [28] focus on K-16 QIST education and conduct workshops and courses suited for various levels.

While the efforts of these organizations and others to support QIST education is important particularly due to the increased career opportunities in both quantum-proficient and quantum-aware/adjacent jobs[3], further efforts to diversify QIST is critical for harnessing the talents of demographic groups that have historically been marginalized in the first quantum revolution. Increasing the visibility and impact of existing underrepresented QIST researchers as well as increasing the number of role models would help with diversification. Changing the culture of disciplines such as physics would help since the inequitable culture in these sub-disciplines can impact the culture of QIST and the extent to which it ends up being diverse, equitable and inclusive as the field matures.

Finally, the interdisciplinary nature of QIST, in which nobody can claim to be an expert, can be an opportunity for diversifying this new field whose culture can be shaped and made more equitable than disciplines such as physics, electrical engineering, or computer science that already have a set culture, which is an impediment to diversity, equity and inclusion. However, the interdisciplinary nature of QIST can also present challenges since student preparation is often siloed into different disciplines, and people with a particular type of training may not feel comfortable being part of an interdisciplinary community in which it is difficult to communicate with people with different types of training. Without adequate support, these issues, e.g., related to sense of belonging [29] in such a community consisting of those with training in physics, engineering, computer science, chemistry, etc., may be particularly challenging for people from historically marginalized groups. Coming up with a shared QIST language that is not heavy on quantum physics jargon (unimportant for many stakeholders with non-physics training) can be useful. Ultimately, forging an empathetic and supportive community culture in which those with different types of training feel comfortable and safe sharing their ideas can go a long way to diversify QIST.

## Acknowledgment

We thank the NSF for award PHY-2309260.